\documentclass[12pt]{iopart}
\usepackage{epsfig}
\begin{document}

\title[Leave-one-out prediction error ...]{Leave-one-out prediction error of systolic arterial pressure time series
 under paced breathing}

\author{N. Ancona\dag\, R. Maestri\P\, D. Marinazzo \ddag\S$\|$, L. Nitti\ddag$\|^+$, M. Pellicoro\ddag\S$\|$, G. D. Pinna\P\ and S.
Stramaglia \ddag\S$\|$
\footnote[3]{To whom correspondence should be addressed
(sebastiano.stramaglia@ba.infn.it)} }

\address{\dag\ Istituto di Studi sui Sistemi Intelligenti per l'Automazione, C.N.R., Bari, Italy}

\address{\ddag\ TIRES-Center
of
Innovative Technologies for Signal Detection and Processing,\\
University of Bari, Italy}

\address{\S\ Dipartimento Interateneo di Fisica, Bari, Italy}

\address{$\|$\ Istituto Nazionale di Fisica Nucleare, Sezione di Bari, Italy}

\address{\P\ Dipartimento di Bioingegneria e
Tecnologie Biomediche, Fondazione Salvatore Maugeri, IRCCS Istituto Scientifico di
Montescano (PV), Italy }

\address{$^+$ Dipartimento di Biochimica Medica, Biologia Medica e Fisica Medica, University of Bari, Italy}

\begin{abstract}
In this paper we show that different physiological states and pathological conditions may
be characterized in terms of predictability of time series signals from the underlying
biological system. In particular we consider systolic arterial pressure time series from
healthy subjects and  Chronic Heart Failure patients, undergoing paced respiration. We
model time series by the regularized least squares approach and quantify predictability
by the leave-one-out error. We find that the entrainment mechanism connected to paced
breath, that renders the arterial blood pressure signal more regular, thus more
predictable, is less effective in patients, and this effect correlates with the
seriousness of the heart failure. The leave-one-out error separates controls from
patients and, when all orders of nonlinearity are taken into account, alive patients from
patients for which cardiac death occurred.
\end{abstract}



\maketitle

\section{Introduction}
Physiological signals derived from humans are extraordinarily complex, as they reflect
ongoing processes involving very complicated regulation mechanisms (Glass 2001), and can
be used to diagnose incipient pathophysiological conditions. Many approaches to
characterization and analysis of physiological signals have been introduced in recent
years, including, for example, studies of: Fourier spectra (Akselrod \etal 1981, Pinna
\etal 2002), chaotic dynamics (Babloyantz \etal 1985, Poon and Merrill 1997), wavelet
analysis (Thurner \etal 1998, Marrone \etal 1999), scaling properties (Nunes Amaral
\etal, 1998, Ashkenazy \etal 2001, Ivanov and Lo 2002), multifractal properties (Ivanov
\etal 1999, Nunes Amaral \etal 2001), correlation integrals (Lehnertz and Elger 1998),
$1/f$ spectra (Peng \etal 1993, Ivanov \etal 2001) and synchronization properties
(Schafer \etal 1998, Tass \etal 1998, Angelini \etal 2004). Less attention has been paid
to the degree of determinism (Kantz and Schreiber 1997) of a physiological time series.
It is the purpose of the present work to show that different physiological states, or
pathological conditions, may be characterized in terms of {\it predictability} of time
series. In particular we consider here {\it predictability} of Systolic Blood Pressure
(SAP) time series under {\it paced respiration}, and show that a suitable index separates
healthy subjects from Chronic Heart Failure (CHF) patients. Systolic blood pressure (SAP)
is the maximal pressure within the cardiovascular system as the heart pumps blood into
the arteries. Paced respiration (breathing is synchronized with some external signal) is
a well-established experimental procedure to regularize and standardize respiratory
activity during autonomic laboratory investigations (Cooke \etal 1998), and a useful tool
for relaxation and for the treatment of chronic pain and insomnia, dental and facial
pain, etc. (Clark and Hirschman 1980, Clark and Hirschman 1990, Freedman and Woodward
1992). Entrainment between heart and respiration rate (cardiorespiratory synchronization)
has been detected in subjects undergoing paced respiration (Schiek \etal 1997, Pomortsev
\etal 1998). Paced breathing can prevent vasovagal syncope during head-up tilt testing
(Jauregui-Renaud \etal 2003); in healthy subjects under paced respiration the
synchronization between the main processes governing cardiovascular system  is stronger
than the synchronization in the case of spontaneous respiration (Prokhorov \etal 2003).
However, a number of important questions remain open about paced breathing, including the
dependence on the frequency of respiration and whether it affects the autonomic balance.
In a healthy cardiorespiratory system, the regime of paced respiration induces {\it
regularization} of related physiological signals (Brown Troy \etal 1993, Pinna \etal
2003), in particular blood pressure time series smoothen and become more deterministic.
To quantify this phenomenon, we face two problems at this point: (i) how may we model the
SAP time series? (ii) what measure of predictability is the most suitable? In the present
paper we model time series by Regularized Least Squares (RLS) approach (Mukherjee \etal
2002). The choice of this class of models is motivated by the fact that it enjoys several
interesting properties. The most important is that such models have high generalization
capacity. This means that they are able to predict complex signals when a finite and
small number of observations of the signal itself are available. Moreover the degree of
nonlinearity present in the modelling, introduced by a kernel method, may be easily
controlled. Finally they allow an easy calculation of the leave-one-out (LOO) error
(Vapnik 1998), the quantity that we use to quantify predictability.   To our knowledge,
this is the first time RLS models are used to model time series; our approach generalizes
the classical autoregressive (AR) approach to time series analysis (Kantz and Schreiber
1997). It is worth mentioning that recently (Shalizi \etal 2004) a measure of
self-organization, rooted on optimal predictors, has been proposed. In the same spirit,
LOO prediction error is related to the degree of organization of the underlying
physiological system.

\section{Method}

\subsection{Regularized least squares linear models for regression}
\label{int} Let us consider a set of $\ell$ independent, identically distributed data
$S=\{ ({\bf x}_i, y_i) \}_{i=1}^\ell$,  where ${\bf x}_i$ is the $n$-dimensional vector
of input variables and $y_i$ is the scalar output variable. Data are drawn from an
unknown probability distribution $p({\bf x},y)$. The problem of learning consists in
providing an estimator $f_{\bf w}:{\bf x}\to y$, out of a class of functions $F({\bf
w})$, called {\it hypothesis space}, parametrized by a vector ${\bf w}$. Let us first
consider the class of linear functions $y={\bf w}\cdot{\bf x}$, where ${\bf w}$ is the
$n$-dimensional vector of parameters. To provide a bias term in the linear function (to
be included if ${\bf x}$ or $y$ have non vanishing mean), a supplementary input variable
(constant and equal to one) is to be included in the input vector. In the regularized
least squares approach, ${\bf w}$ is chosen so as to minimize the following functional:
\begin{equation}\label{lagrangian-function-without-b}
L(\mathbf{w}) = \frac{1}{\ell} \left[ \sum_{i=1}^\ell \left ( y_i - {\bf w}\cdot {\bf
x}_i \right )^2 + \lambda || {\mathbf w} ||^2\right],
\end{equation}
where $|| \mathbf{w} || = \sqrt{\bf{w}\cdot\bf{w}} $ is the Euclidean norm induced by the
scalar product. The first term in functional $L$ is called {\it empirical risk}, the mean
square prediction error evaluated on the training data; the second one ({\it
regularization term}) can be motivated geometrically by the following considerations. Let
us view data $({\bf x}_i, y_i)$ as points in a $(n+1)$-dimensional space. Each function
$y={\bf w}\cdot{\bf x}$ determines an hyperplane in this space, approximating data
points. The prediction square error on point $i$ is $\epsilon_i =(y_i - {\bf w}\cdot{\bf
x}_i)^2$; let $d_i$ be the square distance between the point and the approximating
hyperplane. It is easy to see that (see fig. 1):
\begin{equation}\label{def-of-square-distance-label}
d_i = \frac{\epsilon_i}{1 + || \bf{w} ||^2}.
\end{equation}
This equation shows that the smaller $|| \bf{w} ||^2$, the better the deviation
$\epsilon_i$ approximates the true distance $d_i$. Hence the role of the regularization
term, whose relevance depends on the value of parameter $\lambda$ and penalizes large
values of $|| \bf{w} ||$, is to let the linear estimator be chosen as the hyperplane
minimizing the mean square distance with the data points. It is easy to minimize
functional $L$ and get the optimal hyperplane:
\begin{equation}
\bf{w}=\left ( \bf{A}+ \lambda \bf{I} \right)^{-1} \bf{b}, \label{e2}
\end{equation}
where $\bf{A}$ is the $n\times n$ matrix given by
\begin{equation}
\bf{A}=\sum_{i=1}^\ell \bf{x}_i \bf{x}_i^\top , \label{e3}
\end{equation}
$\bf{b}$ is the $n$-dimensional vector given by
\begin{equation}
{\bf b}=\sum_{i=1}^\ell y_i \bf{x}_i , \label{e4}
\end{equation}
while $\bf{I}$ stands for the identity matrix.

The empirical risk $E_e=1/\ell\sum_{i=1}^\ell \epsilon_i$ is not a good measure of the
quality of the estimator. What matters is the generalization ability, i.e. the prediction
error on data points which have not been used to train the estimator. The following
measure of the generalization performance, known as LOO procedure, is both intuitive and
statistically robust (one can show that LOO error is almost unbiased, see Luntz and
Brailovsky 1969). For each $i$, data point $i$ is removed from the data set. The
approximating hyperplane is then determined on the basis of the residual set of $\ell -1$
points;  the square prediction error by this hyperplane on point $i$ will be denoted
$\epsilon_i^{loo}$. The LOO error is then defined as follows:
$E_{loo}=1/\ell\sum_{i=1}^\ell \epsilon_i^{loo}$. In principle, calculation of $E_{loo}$
requires the estimation of $\ell$ hyperplanes, thus rendering this procedure unfeasible,
or at least unpractical. However the class of models, we are considering here, allows
calculating LOO error after inversion of only one $n\times n$ matrix. It can be shown
(Mukherjee \etal 2002) that:
\begin{equation}
E_{loo}={1\over \ell}\sum_{i=1}^\ell \left({y_i-{\bf w}\cdot{\bf x}_i\over 1 -{\bf
G}_{ii}}\right)^2, \label{e5}
\end{equation}
where ${\bf w}$ is trained on the full data set, using (\ref{e2}), and $\bf{G}$ is an
$\ell\times\ell$ matrix given by
\begin{equation}
\bf{G}=\bf{X}^\top \left ( \bf{A}+ \lambda \bf{I} \right)^{-1} \bf{X};
\end{equation}
here we denote $\bf{X}$ the $n\times \ell$ matrix whose columns are input data
$\{\bf{x}_i\}$.
\begin{figure}[ht!]
\begin{center}
\epsfig{file=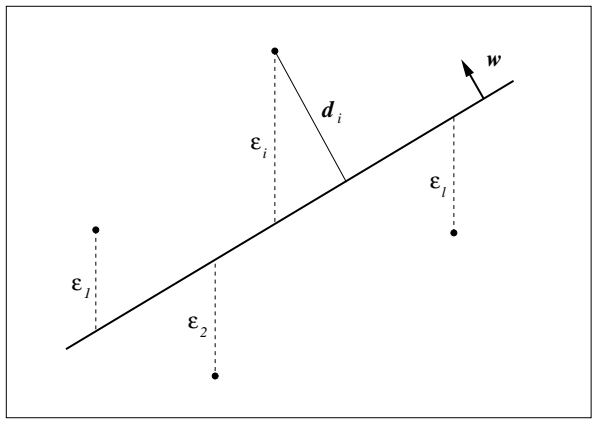,height=5.5cm}
\end{center}
\caption{{\small Geometrical interpretation of regularization \label{fig1}}}
\end{figure}

The value of the parameter $\lambda$ is to be tuned to minimize the LOO error. In other
words, this free parameter is to be tuned to enhance the generalization capability of the
model. It is useful, for the nonlinear extension of these models, to express $\bf{w}$ as
a linear combination of the vectors $\bf{x}_i$ for $i = 1, 2, ..., \ell$. Indeed, if
$\ell
> n$ one can suppose that vectors $\{\bf{x}_i\}$ span all the $n$-dimensional space,
constituting an over-complete system of vectors. This means that there exist $\ell$
coefficients $\bf{c}$ = $(c_1, c_2, ..., c_\ell)^\top$ such that:
\begin{equation}\label{w1}
\bf{w} = \bf{X} \bf{c}.
\end{equation}
Simple calculations yield
\begin{equation}\label{w2}
\bf{c} =  \left (\bf{K} + \lambda \bf{I} \right)^{-1} \bf{y},
\end{equation}
where $\bf{K}=\bf{X}^\top \bf{X}$ is a $\ell \times \ell$ matrix with generic element
$K_{ij} = \bf{x}_i \cdot \bf{x}_j$, whereas $\bf{y}$ = $(y_1, y_2, ..., y_\ell)^\top$ is
a vector formed by the $\ell$ values of the output variable. The prediction $y$, in
correspondence to an input vector $\bf{x}$, may then be written as a sum over input data:
\begin{equation}\label{w3}
f:{\bf x}\to y=\sum_{i=1}^\ell c_i\; \bf{x_i}\cdot \bf{x}.
\end{equation}
Equations (\ref{w2},\ref{w3}) shows that the evaluation of the linear predictor as well
as the computation of the parameter vector $\bf{c}$ involve only scalar products of data
in the input space. This property allows to extend the regularized linear models to the
non linear case, as we describe in the next subsection.

\subsection{Nonlinear models}
The extension to the general case of non linear predictors is done by mapping the input
vectors ${\bf x}$ in a higher dimensional space ${\cal H}$, called {\it feature space},
and looking for a linear predictor in this new space.  Let $\Phi ({\bf x}) \in {\mathcal
H}$ be the image of the point ${\bf x}$ in the feature space, with:

$$
\Phi({\bf x}) = (  \phi_1 ({\bf x}),  \phi_2 ({\bf x}), ...,  \phi_N ({\bf x}), ... )
$$

\noindent where $\{ \phi \}$ are real functions. Note that the number of components of
the feature space can be finite, countable or even infinite uncountable. Moreover,
suppose that one of the features be constant. This hypothesis allows to write the linear
predictor in the feature space ${\cal H}$ without making explicit the bias term. In the
feature space induced by the mapping $\Phi$, a linear predictor takes the form:

\begin{equation}\label{feature-space}
y = f({\bf x}) =  \bf{w} \cdot \Phi ({\bf x})
\end{equation}

\noindent where now $\bf{w}$, according to the nature of the feature space, may have
finite or infinite number of components. Again, we hypothesize that $\bf{w}$ may be
written as a linear combination of the vectors $\Phi ({\bf x}_i)$ with $i=1, 2, ...,
\ell$ (if this hypothesis would not be met, we thus determine a solution, constrained in
the subspace, of the feature space, spanned by vectors $\{\Phi ({\bf
x}_i)\}_{i=1,\ell}$). This means that there exist $\ell$ coefficients $(c_1, c_2, ...,
c_\ell)^\top$ such that:

\begin{equation}\label{w-as-linear-combination-of-phi-x}
{\bf w} = \sum_{i=1}^\ell c_i \Phi ({\bf x}_i).
\end{equation}
In this hypothesis, the linear predictor in the feature space ${\cal H}$ takes the form:

\begin{equation}\label{linear-predictor-in-feature-space}
y = f({\bf x}) = \sum_{i=1}^\ell c_i \Phi ({\bf x}_i)\cdot \Phi ({\bf x}),
\end{equation}
\noindent and, therefore, will be non-linear in the original input variables. The vector
$\bf{c}$ is given by (\ref{w2}) with $\bf{K}$ the $\ell \times \ell$ matrix with generic
element $K_{ij} = \Phi(\bf{x}_i) \cdot \Phi(\bf{x}_j)$. Note that the evaluation of the
predictor on new data points and the definition of the matrix $\bf{K}$ involve the
computation of scalar products between vectors in the feature space, which can be
computationally prohibitive if the number of features is very large. A possible solution
to these problems consists in making the following choice:
$$
\Phi({\bf x}) = (  \sqrt{\alpha_1}\psi_1 ({\bf x}),  \sqrt{\alpha_2}\psi_2 ({\bf x}),
...,\sqrt{\alpha_N} \psi_N ({\bf x}), ... )
$$
where $\alpha_i$ and $\psi_i$  are the eigenvalues and eigenfunctions of an integral
operator whose kernel $K({\bf x},{\bf y})$ is a positive definite symmetric function.
With this choice, the scalar product in the feature space becomes particularly simple
because
\begin{equation}\label{scalar-product}
\Phi({\bf x}_i)\cdot \Phi({\bf x}_j) =  \sum_{\gamma} \alpha_\gamma \psi_\gamma ({\bf
x}_i) \psi_\gamma ({\bf x}_j)= K({\bf x}_i, {\bf x}_j),
\end{equation}
where the last equality comes from the Mercer-Hilbert-Schmidt
theorem for positive definite functions (Riesz and Nagy 1955). The
predictor has, in this case, the form:
\begin{equation}\label{notlinear}
y = f({\bf x}) = \sum_{i=1}^\ell c_i K({\bf x}_i,{\bf x}).
\end{equation}
Analogously the LOO error can be calculated as follows:
\begin{equation}
E_{loo}={1\over \ell}\sum_{i=1}^\ell \left({y_i-\sum_{j=1}^\ell  K_{ij}c_j\over 1 -{\bf
G}_{ii}}\right)^2, \label{e9}
\end{equation}
where the matrix $\bf{G}$ can be shown to be equal to ${\bf K}\left( {\bf K}+\lambda {\bf
I}\right)^{-1}$. Many choices of the kernel function are possible, for example the
polynomial kernel of degree $p$ has the form $K({\bf x},{\bf y})=\left( 1+{\bf
x}\cdot{\bf y}\right)^p$ (the corresponding features are made of all the powers of ${\bf
x}$ up to the $p$-th). The RBF Gaussian kernel is $K({\bf x},{\bf y})=\exp{-\left({||
{\bf x}-{\bf y}||^2/ 2\sigma^2}\right)}$ and deals with all the degrees of nonlinearity
of ${\bf x}$. Specifying the kernel function $K$ one determines the complexity of the
function space within  which we search the predictor, similarly to the effect of
specifying the architecture of a neural network, that is number of layers, number of
units for each layer, type of activation functions which define the set of functions that
the neural network implements. Notice that, depending on the kernel function, we can have
a countable or even an uncountable number of features. The last case apply, for example,
to the Gaussian function. Use of kernel functions to implicitly perform projections, the
{\it kernel trick}, is at the basis of Support Vector Machines, a technique which has
found application in several fields, including Medicine (Bazzani \etal 2001).

\section{Results}
\subsection{Physiological data\label{physiol}}  
\begin{figure}[ht!]
\begin{center}
\epsfig{file=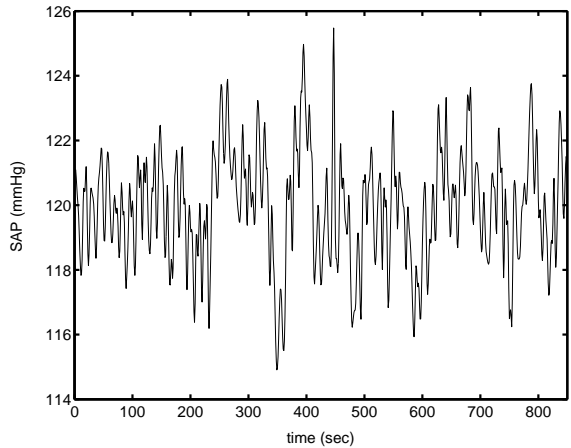,height=6cm}
\end{center}
\caption{{\small The time series  of the systolic arterial pressure for one of the
subjects examined. \label{figsap}}}
\end{figure}

Our data  are from 47 healthy volunteers
(age: $53\pm 8$ years, M/F: 40/7) and 275 patients with chronic heart failure (CHF) (age:
$52\pm 9$ years, LVEF: $28\pm 8 \%$, NYHA class: $2.1\pm 0.7$, M/F: 234/41), caused
mainly by ischemic or idiopathic dilated cardiomyopathy ($48\%$ and $44\%$ respectively),
consecutively referred to the Heart Failure Unit of the Scientific Institute of
Montescano, S. Maugeri Foundation (Italy) for evaluation and treatment of advanced heart
failure. Concerning the second group, cardiac death occurred in 54 ($20\%$) of the
patients during a 3-year follow-up, while the other 221 patients were still alive at the
end of the follow-up period. All the subjects underwent a 10 min supine resting recording
in paced respiration regime (Cooke \etal 1998, Rzeczinski \etal 2002). To perform paced
breathing, subjects were asked to follow a digitally recorded human voice inducing
inspiratory and expiratory phases, at 0.25 Hz frequency. Non invasive recording of
arterial blood pressure at the finger (Finapres device) was performed. For each cardiac
cycle, corresponding values of SAP were computed and re-sampled at a frequency of 2 Hz
using a cubic spline interpolation. As an example In Fig. \ref{figsap} we report the SAP
time series for one of the subjects.

Let us denote $\{{x}_i\}_{i=1,.,N}$ the time series of SAP values, which we assume to be
stationary (this assumption is justified by the short length of the recording). The
models previously introduced are used to make predictions on the time series. We fix the
length of a window $m$, and for $k=1$ to $\ell$ (where $\ell=N-m$), we denote ${\bf
x}_k=(x_{k+m-1}, x_{k+m-2},...,x_{k})$ and $y_k=x_{k+m}$; we treat these quantities as
$\ell$ realizations of the stochastic variables ${\bf x}$ (input variables) and $y$
(output variable). In the preprocessing stage, the time series are normalized to have
zero mean and unit variance, but are not filtered. We use $m=30$, so that the input
pattern receives contributions from frequencies greater than $0.066$ Hz, thus including
part of LF (low frequency $0.04 - 0.15$ Hz) and HF (high frequency $0.15 - 0.45$ Hz)
frequency bands, the major rhythms of heart rate and blood pressure variability. All the
formalism previously described is applied to model the dependency of $y$ from ${\bf x}$,
i.e. to forecast the time series on the basis of $m$ previous values: LOO error is a
robust measure of its predictability. We use Gaussian kernel and polynomial of 1, 2 and 3
degree.

To show the role of the parameter $\lambda$, in fig. 3 we depict, for a typical control
subject, both the LOO error and the empirical error versus $\lambda$. As $\lambda$
increases, the empirical risk monotonically increases, whilst the LOO error shows a
minimum at a finite value of $\lambda$ ensuring the best generalization capability. We
fix the value of $\lambda$ once for all subjects, by minimizing the average LOO error on
a subset made of an equal number of control and CHF time series. This procedure yields
$\lambda =0.01$ for Gaussian kernel and polynomial of 1, 2 degree, whilst for the third
order polynomial kernel the optimal value we find is $\lambda =0.1$. \footnote{For
Gaussian kernel, also $\sigma$ was similarly tuned to minimize LOO error, and fixed equal
$8.5$.}
\begin{figure}[ht!]
\begin{center}
\epsfig{file=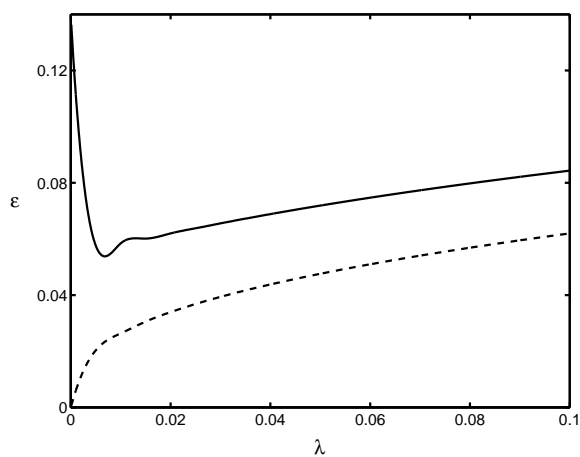,height=6cm}
\end{center}
\caption{{\small For a typical control subject, the LOO error (continuous line) and the
empirical error (dashed line) are represented versus $\lambda$. A Gaussian kernel, with
$\sigma=8.5$, is used. \label{figerr}}}
\end{figure}

We thus evaluate the LOO error for all the 322 subjects (Table 1). In any case, healthy
subjects are characterized by a smaller LOO error than patients. Moreover, dead CHF
patients show greater LOO error than still alive patients. Hence the seriousness of the
heart disease appears to be correlated to the LOO error. The regularized linear model
seems to be the best model of SAP time series. We verify that LOO errors from controls
and patients are Gaussianly distributed and check the homogeneity of the variances of the
two groups; we apply the t-test to evaluate the probability that LOO error values,
relative to controls and patients, were drawn from the same distribution (the null
hypothesis) (table 2). For all kernels, the null hypothesis can be rejected, also after
the Bonferroni correction (which lowers the threshold to $0.05/4=0.0125$). The Gaussian
kernel shows the best separation between the two classes. We have also tested the
separation between dead and alive patients, and the results are also displayed in Table
2. Only when the Gaussian kernel is used the p-value is lower than $0.0125$: since all
orders of nonlinearities contribute to the Gaussian modelling, this result suggests that
the phenomenon here outlined is an effect with strong nonlinear contributions.

\begin{table}[ht!]
\caption{\label{tabe1}Mean values of LOO error.}
\begin{indented}
\item[]\begin{tabular}{@{}lllll}
\br
Kernel&Controls&CHF &CHF alive &CHF dead\\
\mr
Gaussian&0.0386&0.0806&0.0767&0.0968\\
1-poly&0.0019 &0.0158 &0.0131 &0.0272\\
2-poly&0.0022 &0.0842 &0.0745 &0.1242\\
3-poly&0.0082 &0.1493 &0.1484 &0.1526\\

\br
\end{tabular}
\end{indented}
\end{table}

\begin{table}[ht!]
\caption{\label{tabe1}P-values.}
\begin{indented}
\item[]\begin{tabular}{@{}lll}
\br
Kernel&Controls vs CHF &CHF alive vs CHF dead\\
\mr
Gaussian&1.03E-08&0.0088\\
1-poly&0.0011&0.1825\\
2-poly&0.0010&0.1289\\
3-poly&0.0121&0.1429\\

\br
\end{tabular}
\end{indented}
\end{table}

\section{Discussion}
We show that LOO prediction error of physiological time series may usefully be used as a
measure of organization of the underlying regulation mechanisms, and can thus be used to
detect changes of physiological state and pathological conditions. We propose use of RLS
models to time series prediction because they allow fast calculation of the LOO error and
their degree of nonlinearity can be easily controlled. We consider here the SAP time
series in  healthy subjects undergoing paced breath, and in patients with heart disease.
We find that the entrainment mechanism connected to paced breath, that renders the
arterial blood pressure signal more deterministic, thus more predictable, is less
effective in patients, and this effect correlates with the seriousness of the heart
failure; paced breathing conditions seem suitable for diagnostics of a human state. In
our opinion, the LOO error, as a measure of determinism and complexity, is a concept that
has potential application to a wide variety of physiological and clinical time-series
data.

\clearpage
\section*{References}
\begin{harvard}
\item[] Akselrod S, Gordon D, Ubel F A, Shannon D C and Cohen R J 1981 Power spectrum analysis of heart rate fluctuation:
a quantitative probe of beat-to-beat cardiovascular control. 1981 {\it Science} {\bf 213}
220-2
\item[] Angelini L, De Tommaso M, Guido
M, Hu K, Ivanov P C, Marinazzo D, Nardulli G, Nitti L, Pellicoro M, Pierro C and
Stramaglia S 2004 Steady-State Visual Evoked Potentials and Phase Synchronization in
Migraine Patients {\it Phys. Rev. Lett.} {\bf 93} 38103-6
\item Ashkenazy Y, Ivanov P C,
Havlin S, Peng C K, Goldberger A L and Stanley H E 2001 Magnitude and Sign Correlations
in Heartbeat Fluctuations {\it Phys. Rev. Lett.} {\bf 86} 1900-3
\item[] Babloyantz G A, Salazar J M and Nicolis C 1985 Evidence of chaotic dynamics of brain
activity during the sleep cycle {\it Phys. Lett. A} {\bf 111} 152-56
\item[] Bazzani A, Bevilacqua A, Bollini D, Brancaccio R, Campanini R,
Lanconelli N, Riccardi A and Romani D 2001 An SVM classifier to separate false signals
from microcalcifications in digital mammograms {\it Phys. Med. Biol.} {\bf 46} 1651-63
\item[] Brown Troy E, Beightol L A, Koh J and Eckberg L D 1993 Important influence of respiration on human
R-R interval power spectra is largely ignored {\it J. Appl. Physiol.} {\bf 75(5)} 2310-7
\item[] Clark M E and Hirschman R 1990 Effects of paced respiration on anxiety reduction
in a clinical population {\it Biofeedback Self Regul.} {\bf 15} 273-84
\item[] Clark M E and Hirschman R 1980 Effects of paced respiration on affective
responses during dental stress {\it J Dent Res} {\bf 59} 1533-7
\item[] Cooke W K, Cox J F, Diedrich A M, Taylor J A, Beightol L A, Ames IV J E,
Hoag J B, Seidel H and Eckberg LD 1998 Controlled breathing protocols probe human
autonomic cardiovascular rhythms, {\it Am. J. Physiol.} {\bf 274} H709-18
\item[] Freedman R R and Woodward S 1992 Behavioral treatment of menopausal hot
flushes: evaluation by ambulatory monitoring  {\it Am. J. Obstet. Gynecol.} {\bf 67}
436-9
\item[] Glass L 2001 Synchronization and rhythmic processes
in physiology, Nature {\bf 410} 277-84
\item[] Ivanov P C, Nunes Amaral L A, Goldberger A L, Havlin S, Rosenblum M B,
Struzik Z, and Stanley, H. E. (1999) Multifractality in healthy heartbeat dynamics. {\it
Nature} {\bf 399} 461-5
\item[]Ivanov P C, Nunes
Amaral L A, Goldberger A L, Havlin S, Rosenblum M G, Struzik Z and Stanley H E 2001 From
1/f noise to multifractal cascades in heartbeat dynamics {\it Chaos} {\bf 11} 641-52
\item[] Ivanov P C and Lo C C 2002 Stochastic Approaches to Modelling of Physiological
Rhythms, in {\it Modelling Biomedical Signals} eds Nardulli G and Stramaglia S (London:
World Scientific) 28-51
\item[] Jauregui-Renaud K, Marquez MF, Hermosillo AG, Sobrino A, Lara
JL, Kostine A, Cardenas M. 2003 Paced breathing can prevent vasovagal syncope during
head-up tilt testing, Can J Cardiol. 19(6) 698-700.
\item[] Kantz H and Schreiber T 1997 {\it Nonlinear time series analysis} Cambridge University
Press
\item[] Lehnertz K, Elger CE 1998 Can epileptic seizures be predicted? Evidence from nonlinear time
series analyses of brain electrical activity {\it Phys. Rev. Lett.} {\bf 80} 5019-22
\item[] Luntz A and Brailovsky V 1969 On estimation of characters obtained in
statistical procedure of recognition {\it Tecnicheskaya Kibernetica} 3 (in russian)
\item[] Marrone A,  Polosa A D, Scioscia G,
Stramaglia S and Zenzola A 1999 Multiscale analysis of blood pressure signals {\it Phys.
Rev. E} {\bf 60} 1088-91
\item[] Mukherjee S, Rifkin R and Poggio T 2002 Regression and Classification with
Regularization {\it Lectures Notes in Statistics: Nonlinear Estimation and
Classification, Proc. MSRI Workshop}, ed Denison D D, Hansen M H, Holmes C C, Mallick B
and Yu B (Berlin: Springer-Verlag) {\bf 171} 107-24
\item[] Nunes Amaral L A, Ivanov P C, Aoyagi N, Hidaka I, Tomono S, Goldberger A L, Stanley H E and
Yamamoto Y 2001 Behavioral-Independent Features of Complex Heartbeat Dynamics {\it Phys.
Rev. Lett.} {\bf 86} 6026-29
\item[] Nunes Amaral L A, Goldberger A L, Ivanov P C and Stanley H
E 1998 Scale-Independent Measures and Pathologic Cardiac Dynamics {\it Phys. Rev. Lett.}
{\bf 81} 2388-91
\item Nunes Amaral L A, Ivanov P C, Aoyagi N, Hidaka I, Tomono S, Goldberger A L, Stanley H E and
Yamamoto Y 2001 Behavioral-Independent Features of Complex Heartbeat Dynamics {\it Phys.
Rev. Lett.} {\bf 86} 6026-29
\item[] Peng C K, Mietus J, Hausdorff J M, Havlin S, Stanley H E and Goldberger A L
1993 Long-range anticorrelations and non-Gaussian behavior of the heartbeat {\it Phys.
Rev. Lett.} {\bf 70} 1343-46
\item[]Pinna GD, Maestri R, Raczak G, and La Rovere MT 2002
Measuring baroreflex sensitivity from the gain function between arterial pressure and
heart period {\it Clin Sci (Lond)} {\bf 103} 81-8.
\item[] Pinna G D, Gobbi E, Maestri R, Robbi E, Fanfulla F and La Rovere M T 2003 Effect of Paced
Breathing on Cardiovascular Variability Parameters {\it
IEEE-EMBS Asian-Pacifical Conference on Biomedical Engineering}
\item[] Pomortsev A V, Zubakhin A A,
Abdushkevitch V G and Sedunova L F 1998 {\it Proc. XVII Congress
of Physiologists of Russia} ed Kuraev G A (Rostov: Rostov State
University) 316
\item[] Poon C S, Merrill C K 1997 Decrease of cardiac chaos in congestive
heart failure.  {\it Nature} {\bf 389} 492-5
\item[] Riesz F and Nagy B S 1955 {\it Functional Analysis} (New
York: Ungar)
\item[] Prokhorov MD, Ponomarenko VI, Gridnev VI, Bodrov MB, and Bespyatov AB 2003
Synchronization between main rhythmic processes in the human cardiovascular system, Phys.
Rev. {\bf E  68}, 041913-041922
\item[] Rzeczinski S, Janson N B, Balanov A G and McClintock
P V E 2002 Regions of cardiorespiratory synchronization in humans under paced respiration
{\it Phys. Rev. E} {\bf 66} 051909-17
\item[] Schafer C, Rosenblum M G, Abel H H 1998 Heartbeat
synchronized with ventilation {\it Nature} {\bf 392} 239-40
\item[] Schiek M, Drepper F R, Engbert R, Abel H H and Suder K 1997 Transition
between two different Cardiorespiratory Synchronization Regimes
during paced respiration {\it Jahreskongress der DPG, Rostock}
{\bf 76}
\item[] Shalizi C R , Shalizi K L and Haslinger R 2004
Quantifying Self-Organization with Optimal Predictors {\it Phys. Rev. Lett.} {\bf 93}
118701-4
\item[] Tass P, Rosenblum MG , Weule J, Kurths J, Pikovsky A, Volkmann J,
Schnitzler A, Freund H-J, Detection of n:m Phase Locking from Noisy Data: Application to
Magnetoencephalography 1998 Phys. Rev. Lett. {\bf 81} 3291-3294
\item[] Thurner S, Feurstein M C and Teich M C Multiresolution Wavelet Analysis of Heartbeat Intervals
Discriminates Healthy Patients from Those with Cardiac Pathology. 1998 Phys. Rev. Lett.
{\bf 80} 1544-7
\item[]Vapnik V 1998 {\it Statistical Learning
Theory} (New York: John Wiley \& Sons, INC)
\end{harvard}

\end{document}